\documentclass[prb,aps,epsf,showpacs,preprint,preprintnumbers,amsmath,amssymb]{revtex4}
\usepackage{graphicx}
\usepackage{dcolumn}
\usepackage{bm}
\begin{document}
\baselineskip=16pt


\title{Supersolid $^4$He Likely Has Nearly Isotropic Superflow} 
\author{ Wayne M. Saslow$^{1}$} 
\email{wsaslow@tamu.edu}
\author{Shivakumar Jolad$^{1,2}$} 
\affiliation{ $^{1}$Department of Physics, Texas A\&M University, College Station, 
TX 77843}
\affiliation{ $^{2}$Department of Physics, Pennsylvania State University, University Park, PA 16802}
\date{\today}

\begin{abstract}
We extend previous calculations of the zero temperature superfluid fraction $f_s$ (SFF) {\it vs} localization, from the fcc lattice to the experimentally realized (for solid $^4$He) hcp and bcc lattices.  The superfluid velocity is assumed to be a one-body function, and dependent only on the local density, taken to be a sum over sites of gaussians of width $\sigma$.  Localization is defined as $\sigma/d$, with $d$ the nearest-neighbor distance.  As expected, for fcc and bcc lattices the superfluid density tensor is proportional to the unit tensor.  To numerical accuracy of three-places (but no more), the hcp superfluid density tensor is proportional to the unit tensor.  This implies that a larger spread in data on $f_s$, if measured on pure crystals, is unlikely to be due to crystal orientation.  In addition, to three decimal places (but no more) the curves of $f_s$ {\it vs} $\sigma/d$ are the same for both the hcp and fcc cases.  An expected value for the localization gives an $f_{s}$ in reasonable agreement with experiment.  The bcc lattice has a similar curve of $f_s$ {\it vs} $\sigma/d$, but is generally smaller because the lattice is more dilute. 

\end{abstract}

\pacs{67.40.w, 67.80.s}

\maketitle
\section{Introduction}

Theories of superfluidity in solids date to 1969, with the work of Andreev and Lifshitz on the long-wavelength, low-frequency behavior of such systems,\cite{A&L} and to 1970, with the work of Leggett on the non-classical rotational inertia (NCRI).\cite{Leggett1,Leggett2}  In equilibrium, the NCRI fraction (NCRIF) is essentially the superfluid fraction (SFF), denoted $f_s$.\cite{Leggett3}  Rotation of the walls of an annulus causes the many-body wavefunction to develop a boundary condition in the rotating frame wherein a non-integral phase change, associated with backflow, accrues when each particle coordinate is transported around the annulus.  An upper bound was developed for the NCRIF using a phase-function $\phi(\vec{r})$ that included particle correlations only in a dependence on the local particle density $\rho(\vec{r})$: what in the present work we call a one-body phase function.  We assume that the particles making up the solid are bosons. 

Although solid $^4$He takes on the hexagonal close-packed structure, Ref.~\onlinecite{Saslow1}  argued that the results would be similar for a face centered cubic structure, and $f_s$ was then calculated for a density taken to be a sum of gaussians at each lattice site, with lattice constant $a_{fcc}$.  This gave $f_s$ as a function of the localization of the gaussians, taken to vary as $\exp(-r^2/b^2)$.\cite{Saslow1}  Thus $f_s$ was given as a function of $b/a_{fcc}$.  An equivalent notation uses the nearest-neighbor distance $d=a_{fcc}/\sqrt{2}$ and the rms width $\sigma=b/\sqrt{2}$ of the 1d gaussian $\exp(-x^2/b^2)$, so that $f_s$ can also be thought of as a function of $\sigma/d$, a lattice-independent quantity that is relevant when later comparison is made with hcp and bcc lattices.  

Recently, Kim and Chan have observed non-classical rotational inertia (NCRI), suggested by Leggett in 1970, for solid $^4$He in Vycor, porous gold, and in bulk.\cite{KCvycor,KCgold,KCbulk}  Given that the extrapolated values of the $T=0$ NCRIF lie in the same 1-2\% range, despite the different sample porosities, it seems unlikely that it is a surface effect.  Recently NCRI has also been observed in solid para-hydrogen (H$_2$),\cite{KCparaH2} an indication that the effect occurs for all bosons that are sufficiently quantum in nature, as measured, for example, by the (dimensionless) de Boer parameter $\Lambda=\hbar^{2}/[\sigma_{LJ}(m\epsilon)]$, where $\sigma_{LJ}$ is the characteristic length and $\epsilon$ is the characteristic energy associated with the Lennard-Jones 6-12 potential $V=4\epsilon[(\sigma_{LJ}/r)^{12}-(\sigma_{LJ}/r)^{6}]$.  

More recent and better converged calculations based on Ref.\onlinecite{Saslow1} yielded that, for the expected localization ratio from 1976, the superfluid fraction would be about 2\%,\cite{Saslow2} in reasonable agreement with the observations of Kim and Chan.  Therefore in the present work we have calculated $f_s$ for an hcp solid whose density is a sum of gaussians at the lattice sites, again within the context of a one-body phase function.   To characteristic accuracy of three decimal places (but no more), $f_s$ for the hcp solid is isotropic, and to the same degree of accuracy (but no more) the curve of $f_s$ {\it vs} $\sigma/d$ has the same form as for the fcc solid.  On the other hand, to the same accuracy the hcp superfluid response is not literally isotropic, and not literally equivalent to the corresponding fcc lattice.  We also obtain$f_{s}$ for the bcc lattice, whose superfluid density, like that for the fcc lattice, is isotropic. 

These results imply that pure crystals of hcp $^4$He should give, to an accuracy of better than a percent,  the same $f_s$ for all flow directions or, equivalently, for azimuthal flow along a fixed experimental cell within which different crystals are grown.  On the other hand, if the crystal is not perfect, boundary effects may cause the superfluid fraction to change.  This may be the origin of the spread in observed values of the NCRIF.\cite{KCbulk}  

Although the density of real hcp $^4$He is certainly not a sum of gaussians, we expect the result that $f_s$ is isotropic for the hcp lattice is robust.  It is not obvious why $f_s$ should be as very nearly isotropic as it is, although many properties of electronic systems are very similar for the fcc and hcp systems. 

\section{Method}

Ref.\onlinecite{Saslow1} showed that a one-body phase function $\phi$ minimizes the flow energy $E=\frac{1}{2}m\int d\vec{r}\rho(\vec{r})\vec{v}_{s}^{2}$, where $\vec{v}_{s}=\frac{\hbar}{m}\vec{\nabla}\phi$, when the equivalent of the continuity equation is satisfied: 

\begin{equation}
0=\vec{\nabla}\cdot\vec{j}(\vec{r})=\dfrac{\hbar}{m}\vec{\nabla}\cdot\Bigl(\rho(\vec{r})\vec{\nabla}\phi(\vec{r})\Bigr).  
\end{equation}

For a periodic system we take 
\begin{equation}
\phi(\vec{r})=\frac{m}{\hbar}\vec{v}_{0}\cdot\vec{r}+\sum_{G\ne0}\phi_{G}\exp(i\vec{G}\cdot\vec{r}), 
\end{equation}
so
\begin{equation}
\vec{v}_{s}(\vec{r})=\vec{v}_{\,0}+\sum_{G\ne0}\vec{v}_{G}\exp(i\vec{G}\cdot\vec{r})
=\vec{v}_{\,0}+\frac{i\hbar}{m}\sum_{G\ne0}\vec{G}\phi_{G}\exp(i\vec{G}\cdot\vec{r}), 
\end{equation}
where $\vec{v}_{0}$ is considered to be a known imposed average superfluid velocity and the $\vec{G}$'s are reciprocal lattice vectors for the hcp lattice.  To find the unknown flow pattern (equivalent to finding the unknown phases $\phi_{G}$), we use (1) to set up a set of linear equations with $\vec{v}_{0}$ as the source term.  From the phase we can calculate the superfluid velocity and the superfluid fraction. 

The fourier transforms satisfy 
\begin{equation}
\rho(\vec{r})=\sum_{G}\rho_{G}\exp(i\vec{G}\cdot\vec{r}), \qquad
\rho_{G}=V^{-1}\int d\vec{r}\exp(i\vec{G}\cdot\vec{r})\rho(\vec{r}),
\end{equation}
where $V$ is the system volume.  Explicitly, the convolution theorem gives for the fourier components of $\vec{j}(\vec{r})=\rho(\vec{r})\vec{v}_{s}(\vec{r})$ that 
\begin{equation}
\vec{j}_{G}=\sum_{G'}\rho_{G-G'}\vec{v}_{G'}
=\sum_{G'\ne0}\rho_{G-G'}\vec{v}_{G'}+\rho_{G}\vec{v}_{0}.  
\end{equation}

Let 
\begin{equation}
\psi_{G}=\frac{i\hbar}{m}\phi_{G}, 
\end{equation}
so $\vec{v}_{G}=\vec{G}\psi_{G}$.  Then the continuity equation implies, for $\vec{G}\ne\vec{0}$, that 
\begin{equation}
0=G\cdot\vec{j}_{G}=\sum_{G'\ne0}\rho_{G-G'}\vec{G}\cdot\vec{v}_{G'}+\rho_{G}\vec{G}\cdot\vec{v}_{0}
=\sum_{G'\ne0}\rho_{G-G'}\vec{G}\cdot\vec{G'}\psi_{G'}+\rho_{G}\vec{G}\cdot\vec{v}_{0}.
\end{equation}
When the term in $\vec{v}_{0}$ is brought to the other side of the equation, physically this means that, along any $\vec{G}\ne\vec{0}$, the flow due to the fourier components is opposite the average flow $\vec{v}_{0}$.  This induced flow opposite to $\vec{v}_{0}$ also holds for $\vec{G}=\vec{0}$, since typically $f_s<1$. 

The $\psi_{G}$'s that determine the flow pattern can be solved for and then substituted into the equation for the average flow current 
\begin{equation}
\vec{j}_{0}\equiv\overleftrightarrow{{\rho}_{s}}\cdot\vec{v}_{0}
=\rho_{0}\vec{v}_{0}+\sum_{G\ne0}\rho_{-G}\vec{v}_{G}
=\rho_{0}\vec{v}_{0}+\sum_{G\ne0}\rho_{-G}\vec{G}\psi_{G}.
\end{equation}
This completely defines the components of the superfluid density tensor.  We obtain its component along $\vec{v}_{0}$ via 
\begin{equation}
\hat{v}_{0}\cdot\overleftrightarrow{{\rho}_{s}}\cdot\hat{v}_{0}
=\rho_{0}+\sum_{G\ne0}\rho_{-G}\frac{\vec{v}_{0}\cdot\vec{v}_{G}}{v_{0}^{2}}
=\rho_{0}+\sum_{G\ne0}\rho_{-G}\psi_{G}\frac{\vec{v}_{0}\cdot\vec{G}}{v_{0}^{2}}.
\end{equation}
These equations are valid for any $\rho_{G}$.  Sometimes we will refer to 
\begin{equation}
f_s=\dfrac{\hat{v}_{0}\cdot\overleftrightarrow{{\rho}_{s}}\cdot\hat{v}_{0}}{\rho_0}
\end{equation}
as the superfluid fraction, although properly we should refer to 
\begin{equation}
\overleftrightarrow{f_s}=\dfrac{\overleftrightarrow{{\rho}_{s}}}{\rho_0}
\end{equation}
as the superfluid fraction tensor. 

We represent a Gaussian in three-space, normalized to unity, by 
\begin{equation}
\rho(\vec{r})=(\frac{1}{\sqrt{\pi}b})^{3}\exp{[-(\frac{r}{b})^{2}]}. 
\end{equation}
In 1d this has rms width $\sigma=b/\sqrt{2}$.  As already noted, Refs.\onlinecite{Saslow1} and \onlinecite{Saslow2}, for fcc lattices plot $f_s$ as a function of $b/a_{fcc}=b/(d\sqrt{2})=\sigma/d$.  In the present work we also consider the hcp lattice and the bcc lattice. 

Let $\rho_{0}=1/V_{0}$ be the number density in terms of the unit cell volume $V_{0}$.  Then for the sum-of-gaussians model the fourier transform of the total density is given by 
\begin{equation}
\rho_{G}=\rho_{0}\exp(-|\vec{G}|^{2}b^{2}/4)\cos(\vec{G}\cdot\vec{u}).
\end{equation}

\section{HCP Lattice}
\subsection{\bf Lattice}
For the hcp lattice we take the hcp lattice constant in the $x$-$y$ plane to be given by $a_{hcp}$, so the nearest neighbor distance $d=a_{hcp}$.  Thus $\sigma/a_{hcp}=\sigma/d$.  The volume of a unit cell is $V_{0,hcp}=\sqrt{2}a_{hcp}^3=\sqrt{2}d^3$, as can be determined from $V_{0,hcp}=\vec{a}_1\cdot\vec{a}_2\times\vec{a}_3$, with the basis set vectors given below.  This is the same as for the fcc lattice with the same nearest-neighbor distance $d=a_{fcc}/\sqrt{2}$, since $V_{0,fcc}=a_{fcc}^3/4=\sqrt{2}d^3$.  

The hcp real-space basis is taken to be (omitting the subscript on $a$) 
\begin{equation}
\vec{a}_{1}=a(1,0,0), \quad\vec{a}_{2}=\frac{a}{2}(1,\sqrt{3},0), \quad\vec{a}_{3}=a\sqrt{\frac{8}{3}}(0,0,1).
\end{equation}
The lattice sites are specified by $R=i\vec{a}_{1}+j\vec{a}_{2}+k\vec{a}_{3}$, where $i,j,k$ are integers. 

Associated with this is an atomic basis set at $\pm\vec{u}$, where 
\begin{equation}
\vec{u}=\frac{a}{4}(1,-\frac{1}{\sqrt{3}},\sqrt{\frac{8}{3}}). 
\end{equation}
The atoms are located at $\vec{R}\pm\vec{u}$, where $2\vec{u}$ has length $d=a$.   This choice is not unique; we could also have employed $\vec{u}=\frac{a}{4}(1,+\frac{1}{\sqrt{3}},\sqrt{\frac{8}{3}})$. 

The reciprocal space basis is given by
\begin{equation}
\vec{b}_{1}=\frac{2\pi}{a}(1,-\frac{1}{\sqrt{3}},0), \quad\vec{b}_{2}=\frac{2\pi}{a}(0,\frac{1}{\sqrt{3}},0), \quad\vec{b}_{3}=\frac{2\pi}{a}(0,0,\sqrt{\frac{3}{8}}).
\end{equation}
Reciprocal lattice vectors (RLVs) are specified by $\vec{G}=m\vec{b}_{1}+n\vec{b}_{2}+p\vec{b}_{3}$, where $m,n,p$ are integers. 

\subsection{\bf Results} 
In order to obtain numerical results one must truncate the sum over RLVs.  To test for convergence we computed $f_s$ as the RLV basis set was increased.  We employed two methods: 

(1) With $m,n,p$ all running from $-N$ to $+N$, we computed $f_s$ {\it vs} $N$ for each value of $b/d=\sqrt{2}\sigma/d$, and each of the flow directions considered, which included (1,0,0), (0,1,0), (0,0,1), (1,1,1), and (1,$-$1,0).  The results for $b/d=0.35$ ($\sigma/d=0.2474$) and (100) and (001) are given in columns two and three of Table I.  The largest value taken for any $b/d$ value was $N=8$.  

(2) Because the RLV length for the $z$ direction is relatively short, to get a better representation of flow along $z$ we often included more fourier components in that direction, so that $p$ ran from $-2N$ to $2N$.  This led to improved convergence rates for $f_s$, denoted by $f'_s$ in columns four and five of Table I.  For this method the largest value taken for any $b/d$ value was $N=7$.  

\begin{table}[htdp]
\caption{Superfluid fraction $f_s$ for $b/d=0.35$ ($\sigma/d=0.2474$) and various flow directions.  For the hcp lattice, columns two and three are by method (1) and columns four and five by method (2), as discussed in the text.  For the fcc lattice, columns six and seven are for two different flow directions, where $\alpha\equiv(111)/\sqrt{3}$.} 
\begin{center}
\begin{tabular}{|c|c|c||c|c|||c|c|}
\hline
	&hcp(1)		&hcp(1)		&hcp(2)		&hcp(2)	&fcc			&fcc\\
\hline
$N$	&$f_{s,xx}$ 	&$f_{s,zz}$ 	&$f'_{s,yy}$ 	&$f'_{s,zz}$&$f_{s,xx}$	&$f_{s,\alpha\alpha}$\\
\hline
2	&0.91992224	&0.923652853	&0.919241677	&0.919773502&0.91950571	&0.919721444\\
\hline
3	&0.91910979	&0.920556418	&0.919038558	&0.919612587&0.91926078	&0.919278107\\
\hline
4	&0.91903729	&0.919752869	&0.919024750	&0.919607992&0.91924305	&0.919244375\\
\hline
5	&0.91902573	&0.919630510	&0.919023744	&0.919607848&0.91924173	&0.919241833\\
\hline
6	&0.91902397	&0.919611569	&0.919023668	&0.919607842&0.91924163	&0.919241638\\
\hline

\end{tabular}
\end{center}
\label{table1}
\end{table}%

Comparing the (most converged) $N=6$ rows of Table I, we conclude that, although in principle $f_s$ is a tensor quantity with $f_{s,xx}=f_{s,yy}\ne f_{s,zz}$, for the hcp lattice it may be treated as a scalar.  Moreover, to the same accuracy, $f_s$ for the hcp and fcc case (columns six and seven of Table II) are indistinguishable (actually, we find $f^{hcp}_{s,xx}<f^{fcc}<f^{hcp}_{s,zz}$), and thus the form of $f_s$ {\it vs} $\sigma/d$ is the same as given by Ref.\onlinecite{Saslow2}, where $(b/a)_{fcc}=\sigma/d$.  For completeness, we present $f_s$ {\it vs} $\sigma/d$ in Figure ~1, along with data for the bcc lattice, which is discussed in the next section. 

\begin{figure}
\vspace{-.2in}
\centerline {
\includegraphics[width=6.2in]{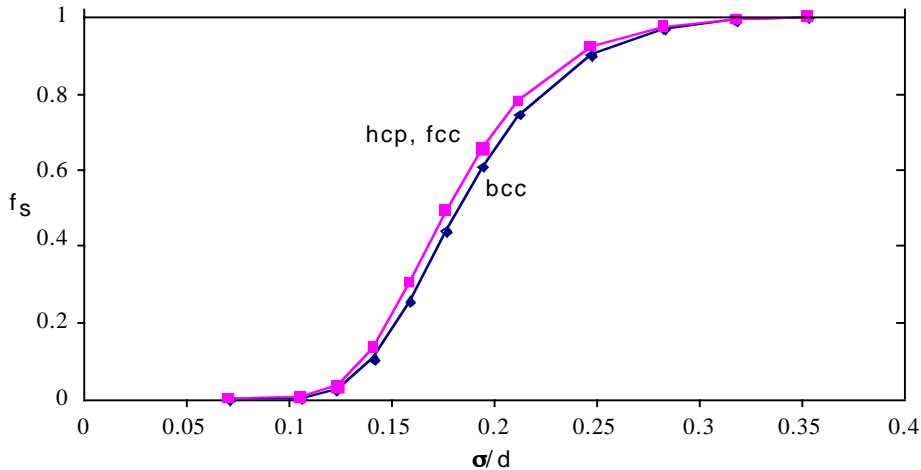}
\vspace{-5.2in}}
\caption{Superfluid fraction $vs\ \sigma/d$ for the fcc, hcp, and bcc lattices.}
\label{fig3}
\end{figure}

To show the extent that $f_s$ is not isotropic, we present Table II for $f_{s,zz}-f_{s,xx}$, which is typically down from $f_{s,zz}$ by three orders of magnitude.  The smaller the value of $\sigma/d$, the less converged the results. 

\begin{table}[htdp]
\caption{Anisotropy $f_{s,zz}-f_{s,xx}$ for hcp lattice for various values of $b/d$.  Also given is the largest value of $N$ needed to obtain convergence (when possible), using method (2) (see text).}
\begin{center}
\begin{tabular}{|c|c|c|c|c|c|c|c|c|c|c|c|l}

\hline
$b/d$ 					&0.70	&0.65	&0.60	&0.55	&0.50	&0.45	&0.40	&0.35	&0.30	&0.25	&0.20\\
\hline
$(f_{s,zz}-f_{s,xx})\times10^3$	&0.001	&0.004	&0.019	&0.062	&0.159	&0.321	&0.495	&0.583	&0.517	&0.330	&0.136\\
\hline
$N$						&2		&3		&3		&3		&3		&3		&4			&6		&7		&7		&7\\
\hline

\end{tabular}
\end{center}
\label{table3}
\end{table}%

For converged values of $f_s$, we have verified that, for flow along, e.g., (111), the current can be obtained by application of $\vec{j}_{0}=\overleftrightarrow{\rho}_{s}\cdot\vec{v}_{0}$ and the tensor components $\rho_{s,xx}$ and $\rho_{s,zz}$ obtained from flow along (100) and (001). 

We have also employed (8) to obtain the three components of the current flow, for various directions of average superflow.  These results are consistent with our studies indicating that the superfluid density tensor is nearly isotropic.  Moreover, it is isotropic in the $x$-$y$ plane when the calculations are performed to convergence, just as the full superfluid tensor density is isotropic for the fcc lattice. 

\section{bcc lattice}

Part of the phase diagram for solid $^4$He, at higher temperatures and near the melting line, includes a bcc phase.  Given that the transition temperature for superflow in solids is well below 0.5~K, whereas the bcc phase does not occur until temperatures near 1.5~K, it is unlikely that the bcc phase will exhibit superflow.  Nevertheless, calculation of $f_s$ should be physically revealing.  We anticipate that for a given $\sigma/d$, since this much more open lattice has a lower density and more space, on average, between sites, that the superfluid fraction typically will be lower than for the corresponding fcc and hcp cases.  Calculations on $f_s$ for this lattice, which has an isotropic superfluid density, bear this out.

\subsection{\bf Lattice and Results}
For the bcc lattice, the real-space basis is taken to be 
\begin{equation}\vec{a}_{1}=(a/2)(-1,1,1), \quad \vec{a}_{2}=(a/2)(1,-1,1), \quad \vec{a}_{3}=(a/2)(1,1,-1).
\end{equation}  
Then $\vec{R}=i\vec{a}_{1}+j\vec{a}_{2}+k\vec{a}_{3}$ gives the lattice sites. 
The reciprocal space basis set is given by 
\begin{equation}
\vec{b}_{1}=(2\pi/a)(0,1,1), \quad \vec{b}_{2}=(2\pi/a)(1,0,1), \quad \vec{b}_{3}=(2\pi/a)(1,1,0).
\end{equation}  
Then the RLVs are $\vec{G}=m\vec{b}_{1}+n\vec{b}_{2}+n\vec{b}_{3}$.  The bcc unit volume is $V_{0,bcc}=a_{bcc}^3/2$, so $\rho_{0,bcc}=2/a_{bcc}^3$.  Moreover, $d=a_{bcc}\sqrt{3}/2$, so  $\rho_{0,bcc}=3\sqrt{3}/4d^3=1.299/d^3$.  Both the fcc and hcp lattices have the higher densities $\rho_{0,fcc}=\rho_{0,hcp}=\sqrt{2}/d^3=1.4142/d^3$.  Table III shows some of our results that are to be compared with Table I for the hcp and fcc cases.  Note the convergence rate and note that for the bcc lattice to have the same value of $f_{s}$ as for the hcp and fcc lattices, $\sigma/d$ must be larger. 

\begin{table}[htdp]
\caption{Superfluid fraction for bcc lattice with $\sigma/d=0.35$ ($b/d=0.4950$) and two flow directions.}
\begin{center}
\begin{tabular}{|c|c|c|}

\hline
$N$	&$f_s$(100)	&$f_s$(111)\\
\hline
2	&0.90084901	&0.90059681\\
\hline
3	&0.90043751	&0.90041712\\
\hline
4	&0.90040634	&0.90040469\\
\hline
5	&0.90040389	&0.90040375\\
\hline
6	&0.90040369	& 0.90040368\\
\hline

\end{tabular}
\end{center}
\label{table4}
\end{table}%

Figure 1 shows the superfluid fraction {\it vs} $\sigma/d$ for the bcc lattice.  It is typically less than that for the fcc and bcc lattices, as expected. 

\section{Conclusions}

Given that Kim and Chan obtain SFF's in the range of 1-2\%, it would appear that the value of $b/d$ is about 0.12.\cite{Saslow2}  However, this value is just in the range where the curve of SFF {\it vs} $b/d$ develops a very steep slope.  If a microscopic calculation of $b/d$ (or, better, of $\rho(\vec{r})$) is accurate to only 10\%, then the actual value of $b/d$ could be overestimated, giving a value for $f_{s}$ perhaps as large as 20\%.  If this is the case, the difference between theory and experiment could possibly be attributed to the error in the microscopic calculations of  $\rho(\vec{r})$, rather than to an inadequacy of the one-body theory of superflow.  However, current microscopic $\rho(\vec{r})$'s are expected to be accurate to a few percent, which should enable a judgement to be made on whether the one-body phase gives an accurate description of the superfluid.  Work is in progress to go beyond a one-body phase.\cite{GalliSaslow}

Finally, we note that, although $f_s<1$, even at temperature $T=0$, this does not imply that thermal excitations take up the slack, since we expect that the normal fluid fraction $f_n=0$ at $T=0$.  Rather, a lattice fraction $f_L$ must be added to Andreev and Lifshitz's macroscopic theory of superflow.\cite{Saslow2}  

\section{Acknowledgements}
WMS would like to acknowledge conversations and correspondence with D. Galli and M. Chan.  This work was supported in part by the Department of Energy, through DOE Grant No. DE-FG03-96ER45598.

\end{document}